\documentclass{article}
\usepackage{arxiv}
\usepackage{amsmath,amsfonts}
\usepackage{algorithmic}
\usepackage{algorithm}
\usepackage{array}
\usepackage[caption=false,font=normalsize,labelfont=sf,textfont=sf]{subfig}
\usepackage{textcomp}
\usepackage{hyperref}
\usepackage{dblfloatfix}
\usepackage{url}
\usepackage{verbatim}
\usepackage{graphicx}
\usepackage{cite}
\usepackage{breqn}
\usepackage{setspace}
\usepackage{geometry}
\usepackage[flushleft]{threeparttable}
\usepackage{graphicx}
\usepackage{tikz}

\newcommand{\subfigimg}[3][,]{%
	\setbox1=\hbox{\includegraphics[#1]{#3}}% Store image in box
	\leavevmode\rlap{\usebox1}% Print image
	\rlap{\hspace*{0pt}\raisebox{\dimexpr\ht1-1\baselineskip}{#2}}% Print label
	\phantom{\usebox1}% Insert appropriate spcing
}

\title{Semi-blind-trace algorithm for self-supervised attenuation of trace-wise coherent noise}

\author{
	Mohammad Mahdi Abedi\textsuperscript{1},
	David Pardo\textsuperscript{2,1,3},
	Tariq Alkhalifah\textsuperscript{4}\\
	\textsuperscript{1}Basque Center for Applied Mathematics, Bilbao, Spain\\
	\textsuperscript{2}University of the Basque Country, Department of Mathematics, Spain\\
	\textsuperscript{3}Ikerbasque, Basque Foundation for Science, Bilbao, Spain\\
	\textsuperscript{4}King Abdullah University of Science and Technology, Thuwal 23955-6900, Saudi Arabia\\
	\textit{Emails:} mabedi@bcamath.org, david.pardo@ehu.es, tariq.alkhalifah@kaust.edu.sa
}

\begin{document}
	\bibliographystyle{unsrt}
	\maketitle
   \footnotetext{Under revision}
	
	\begin{abstract}
		Trace-wise noise is a type of noise often seen in seismic data, which is characterized by vertical coherency and horizontal incoherency. Using self-supervised deep learning to attenuate this type of noise, the conventional blind-trace deep learning trains a network to blindly reconstruct each trace in the data from its surrounding traces; it attenuates isolated trace-wise noise but causes signal leakage in clean and noisy traces and reconstruction errors next to each noisy trace. To reduce signal leakage and improve denoising, we propose a new loss function and masking procedure in semi-blind-trace deep learning. Our hybrid loss function has weighted active zones that cover masked and non-masked traces. Therefore, the network is not blinded to clean traces during their reconstruction. During training, we dynamically change the masks' characteristics. The goal is to train the network to learn the characteristics of the signal instead of noise. The proposed algorithm enables the designed U-net to detect and attenuate trace-wise noise without having prior information about the noise. A new hyperparameter of our method is the relative weight between the masked and non-masked traces' contribution to the loss function. Numerical experiments show that selecting a small value for this parameter is enough to significantly decrease signal leakage. The proposed algorithm is tested on synthetic and real off-shore and land datasets with different noises. The results show the superb ability of the method to attenuate trace-wise noise while preserving other events. An implementation of the proposed algorithm as a Python code is also made available.
		
		
	\end{abstract}
	
	\section{Introduction}
	Recorded  seismic data include different types of noise that can cover and interfere with the objective signals. Due to the band-limited nature of the seismic recordings, seismic noises are not genuinely random and usually have coherencies in one or more directions. One type of noise we address in this study is coherent along the recorded traces (vertical direction) and incoherent in the offset (horizontal) direction. This trace-wise noise includes cross-feed noise (after stacking), swell noise, streamers bird noise, velocity sensor noise, and spike-like noise in marine data (see \cite{hlebnikov2021}), receivers malfunctioning noise, and noises resulting from external sources of energy in common receiver gathers (see \cite{strobbia2022}). Also, in pseudo deblending, the blending noises become trace-wise incoherent after specific rearrangements of data \cite{wang2021}.
	
	Many seismic denoising methods rely on the predictability of signals using their coherency, presumed geometrical shape, or other characteristics. Methods that use signal prediction algorithms include the f-x prediction filter (e.g., \cite{canales1984,spitz1991}), plane-wave destruction (\cite{fomel2002}), polynomial fitting (\cite{liu2011}), and dip steering (e.g., \cite{huo2017}). Methods that filter noise based on the coherency (i.e. smoothness) of the signal include local mean (e.g., \cite{claerbout1985}), median (e.g., \cite{zhu2004}), and nonlocal means (e.g., \cite{bonar2012}) filtering. On the other hand, methods that try to separate signal and noise in an alternative domain use Fourier (e.g., \cite{Stewart1989, hashemi2008}), wavelet (e.g., \cite{mousavi2016hybrid,mehr2017}), curvelet (e.g., \cite{neelamani2008}), seislet (e.g., \cite{fomel2010}), Radon (e.g., \cite{trad2003}), and other transforms. Recent data-driven methods for noise attenuation include dictionary learning (e.g., \cite{nazari2017,wang2019adaptive,zhou2020,almadani2021dictionary, sui2023}), deep learning (e.g., \cite{Yu2019, zhao2019,zhu2019seismic, saad2020,saad2021fully, wang2022learning, farmani2023,markovic2023diffraction}), and hybrid methods (e.g., \cite{farmani2022,Qian2022,liu2023dl2}).  The deep learning methods either directly attenuate the noise, or detect the noise for other noise attenuation algorithms.
	
	Many deep learning applications employ supervised learning methods. For denoising purposes, supervised methods require pairs of clean and noisy data samples that are often unavailable for real seismic applications. Supervised methods use synthetic labeled data for training, therefore, they often fail to understand the reality of field data features (\cite{alkhalifah2022mlreal}). In the absence of pretrained networks and a large number of labeled samples, self-supervised methods provide alternate learning using the target data itself. Self-supervised denoising methods are trained and used on noisy data without having a clean version available. Self-supervised learning has the disadvantage that the training should be performed for each objective dataset. However, such target-oriented training is usually performed on small datasets and for a few epochs. In this way, self-supervised methods also avoid the computationally intensive pretraining on comprehensively representative large datasets and the need for acquiring labels for supervised learning. 
	
	Blind-spot denoising (also named noise to void) is a self-supervised algorithm that removes noise using the predictability of signals and unpredictability of noise within the receptive field of each data point (\cite{krull2019}). This method assumes that the noise is pixel-wise independent, therefore, it is suitable for attenuation of incoherent noise. An extension of this method (named structured noise to void) aims to attenuate noise along extended blind masks (\cite{broaddus2020}), in which the noise is deemed coherent. In seismic studies, adaptations of these methods are used for the attenuation of random noise (\cite{meng2021, birnie2021}), trace-wise coherent noise (\cite{birnie2022transfer,liu2022coherent,abedi2023multi}), and deblending (\cite{wang2022}). 
	
	A key requirement for denoising methods is to preserve the signal quality while addressing the denoising task \cite{ovcharenko2020a}. In blind-trace denoising \cite{liu2022coherent}, all traces --whether noisy or not-- are reconstructed from the neighboring traces. This results in signal leakage in the entire data, and reconstruction errors close to the noisy traces. To overcome these problems, we propose a semi-blind-trace deep learning algorithm that differentiates between the noisy and clean data during the reconstruction, using a new masking procedure and a new loss function. The proposed algorithm simultaneously performs noise detection and attenuation. Thus, the contribution of this paper can be summarized in the following:
	
	\begin{itemize}
		\item We propose a new masking and loss measurement procedure to enhance blind-trace deep learning.
		\item We show the improved denoising capabilities and data fidelity using the proposed semi-blind-trace deep learning.
		\item We explain how to define the new hyperparameters for an optimal result.
	\end{itemize}
	
	We first explain blind-trace denoising for self-supervised attenuation of trace-wise coherent noise in seismic data and explain the reason behind the signal leakage and poor denoising performance in the vicinity of the noisy traces. Then, we propose our semi-blind-trace deep learning methodology, describe the implementation of the algorithm, study the effect of its hyperparameters,  and show two applications on field seismic data.
	
\graphicspath{{./Figures/}}

	\begin{figure*}
		\centering
		\begin{tabular}{@{}p{\linewidth}@{\quad}p{\linewidth}@{}}
			\centering
			\subfigimg[width=.8\linewidth]{(a)}{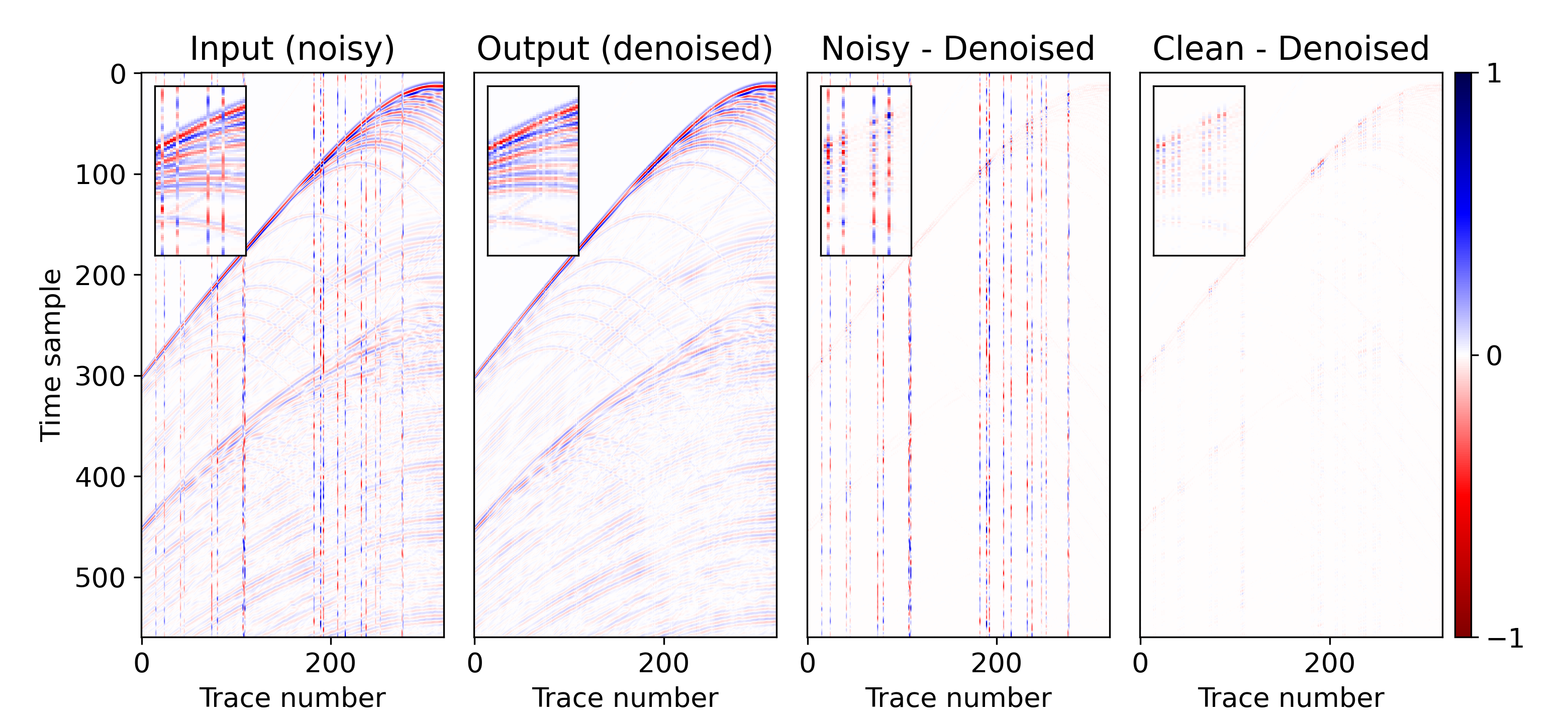} 
			\\
			\subfigimg[width=.45\linewidth]{(b)}{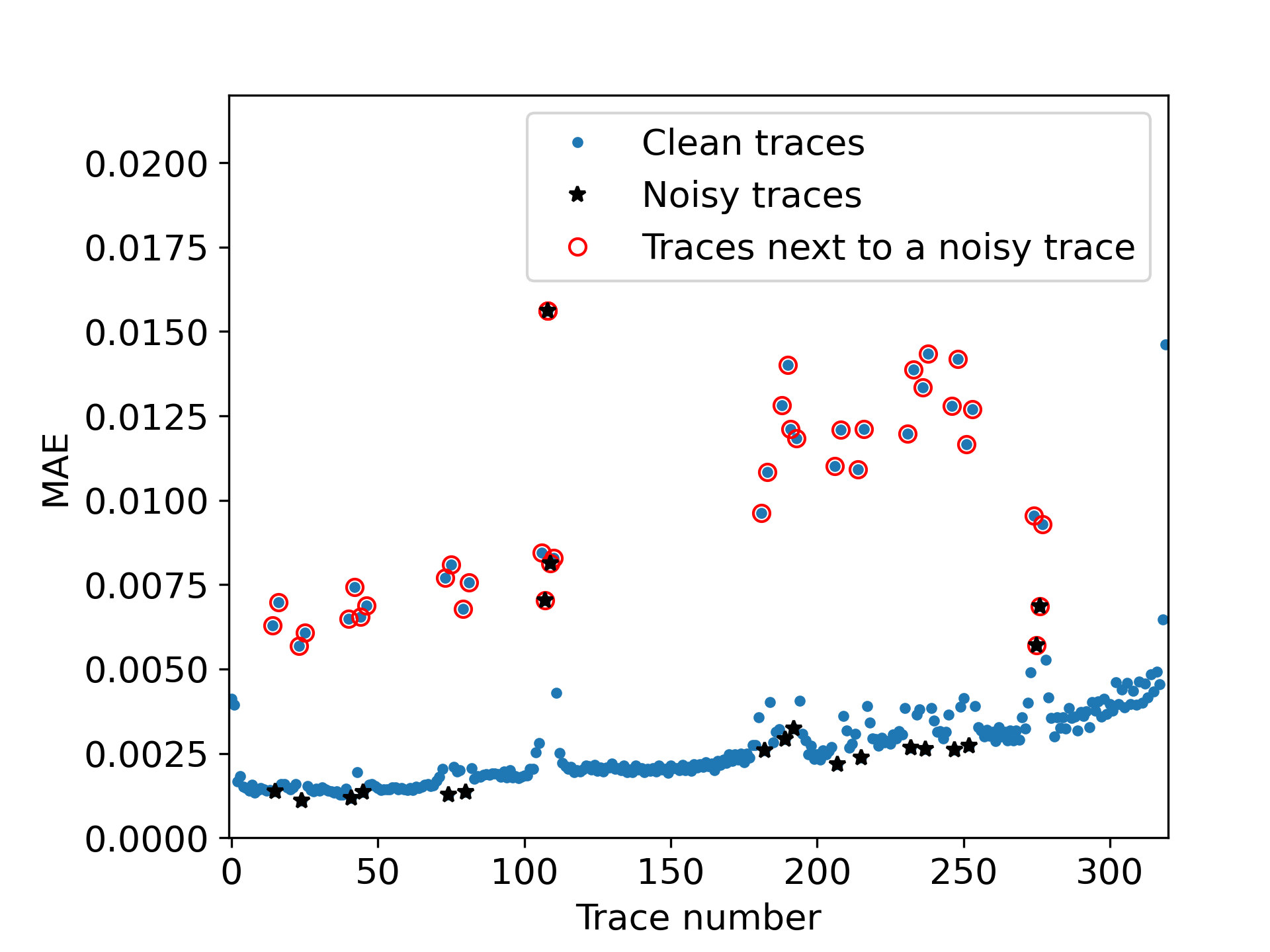} 
			\subfigimg[width=.45\linewidth]{(c)}{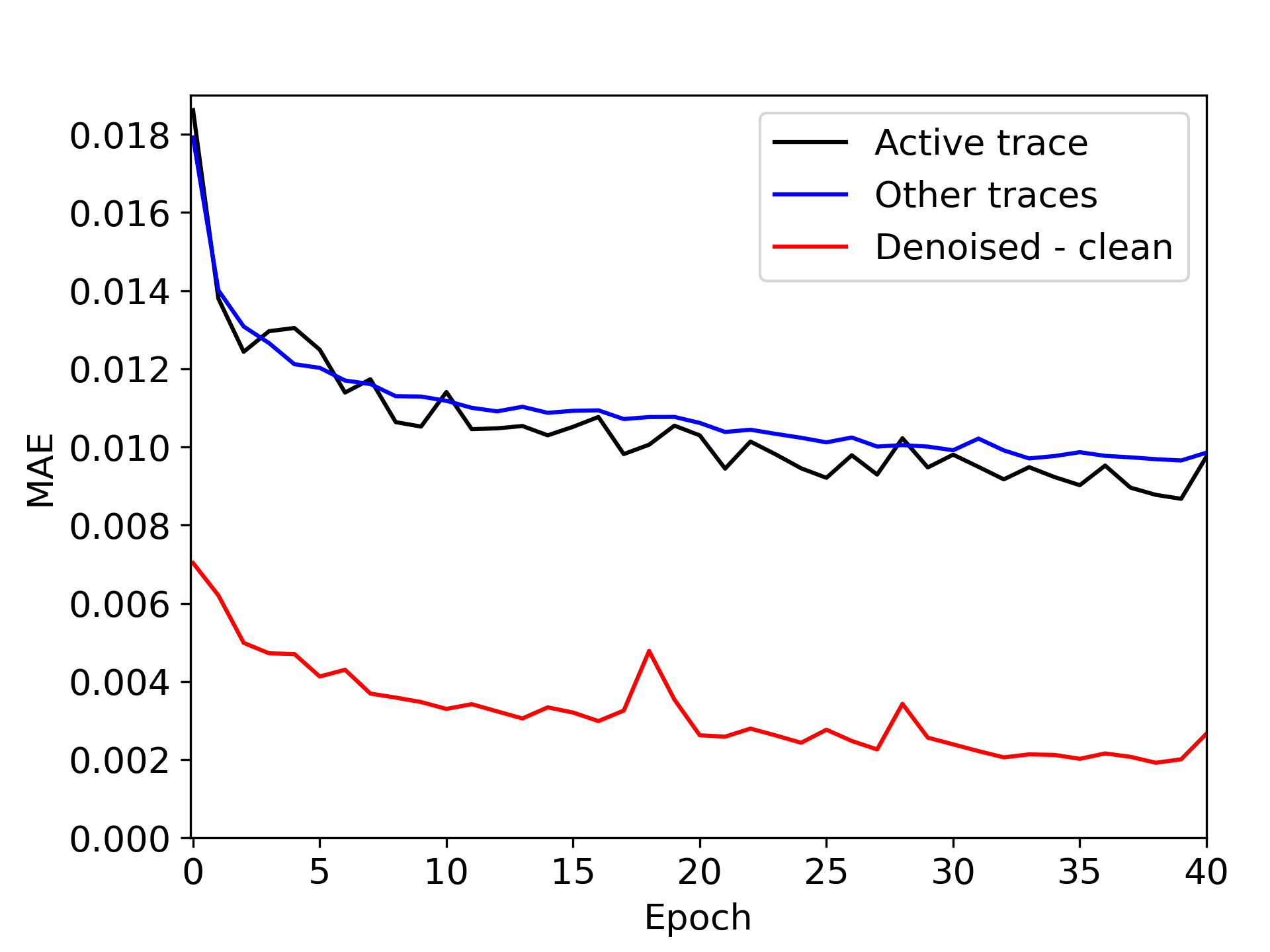}
		\end{tabular}
		\caption{Application of the conventional blind-trace denoising on synthetic data. a) From left to right: the input noisy data, the self-supervised denoised version, the removed noise, and the signal leakage. Overlapping windows show a zoomed portion. b) The average error of each trace after denoising. Red circles mark the (clean or noisy) traces neighboring a noisy trace.  c) Evolution of the loss, and two metrics. Loss (in black) is the mean difference between the DNN output and the noisy data at the active trace. The first metric (in blue) is calculated similarly but for nonactive traces. The second metric (in red) is the difference between the DNN output and the clean data.}
		\label{fig1}
	\end{figure*}

	\section{Blind-trace denoising}
	In the structured noise-to-void method \cite{broaddus2020}, a part of each data patch (whose shape is adapted to the structure of the target noise in the receptive field of each data point) is replaced with random noise, and then a DNN is trained to reconstruct that part. The process of replacing parts of the data with noise before feeding the data to a DNN is called ''masking''. In its seismic adaption  \cite{wang2022,liu2022coherent,abedi2023multi}, randomly selected traces in the input data are masked and then a DNN is trained to reconstruct them. Since the network is blinded to a trace in the data, the method is called ''blind-trace denoising''. Several masked versions of each data sample are created using randomly selected traces for masking. A network is forced to reconstruct each trace from its surrounding traces (ignoring the data or noise in the trace itself). The loss function is defined as the mean absolute error (MAE) between the original and the reconstructed masked traces:
	\begin{dmath}
		\label{eq1}
		L = \frac{1}{ \sum \textbf{M}} \sum \textbf{M} \odot \bigg |\textbf{D} - f (\textbf{D}_m)\bigg |,
	\end{dmath}
	\noindent
	where  $ \textbf{D} $ is the original data matrix, $\textbf{D}_m$ is the masked data, $f(\textbf{D}_m )$ is the DNN output for $\textbf{D}_m$,  $ \textbf{M} $ is the mask matrix with the same dimensions as $\textbf{D}$, and  $\odot$ is the elementwise multiplication symbol.  $ \textbf{M} $ consists of ones over the randomly selected traces and zeros elsewhere. Its role in the loss function is to limit the calculation of loss to the masked traces. A trace that contributes to the loss is also called an ''active trace''.
	
	\begin{figure*}[]
		\centering
		\includegraphics[width=.9\linewidth]{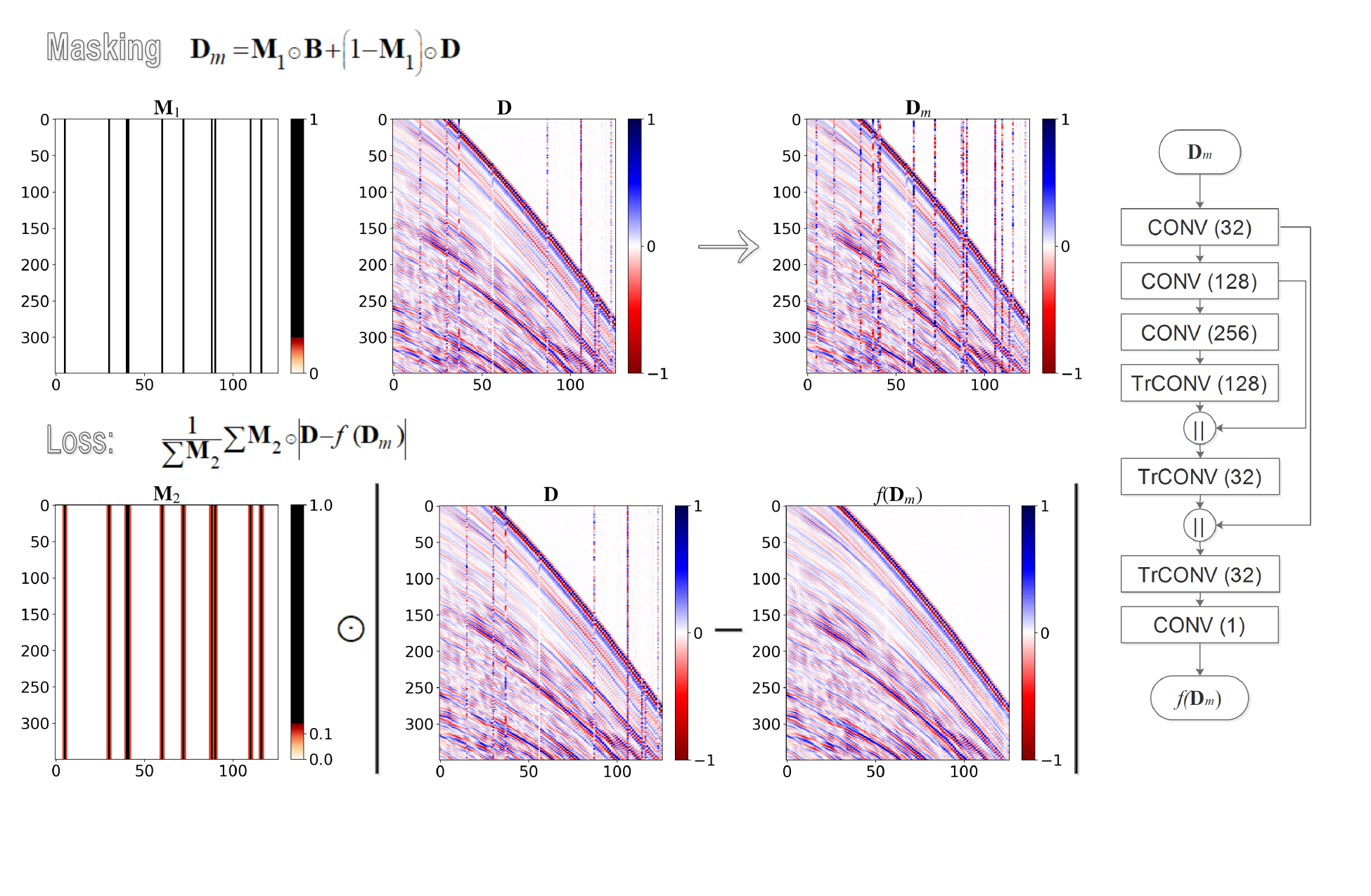}
		\caption{The process of masking (upper panels) and calculation of loss (lower panels), besides the designed architecture (on the right) for the proposed semi-blind-trace algorithm. For each input $ \textbf{D} $, we randomly generate several instances of $ \textbf{M}_1 $ for masking the network's input, and $ \textbf{M}_2 $ to define active traces in the loss function, following the $ \textbf{M}_1 $ pattern. $ \textbf{B} $  is band-limited random noise. In the architecture, the $"||"$ symbol implies concatenation. Details of the network are in Table \ref{tab:table1}. The network is trained to reproduce $ \textbf{D} $ from $  \textbf{D}_m $, but produces its denoised version ($ f( \textbf{D}_m) $).}
		\label{fig2}
	\end{figure*}

	A DNN is trained using different masked samples as its input and the original (noisy) sample as the label. During training, the network is expected to learn to reconstruct each trace from the neighboring traces in which the coherency of the signal is utilized for the reconstruction. Since the noise is only coherent along the trace, the reconstruction process is expected to fail to learn the noise. Therefore, although the network is trained to reproduce the original noisy data, it initially produces a denoised version of it in a self-supervised fashion.
	
	To test the conventional blind-trace denoising, we use 200 shot gathers from the 2004 BP benchmark 2D synthetic data. A number between 10 to 20 traces in each shot gather is contaminated with bandlimited coherent noise to form a dataset of 200 noisy samples. The shot interval is 50 m, the receiver interval is 12.5 m, and the time sampling interval is 6 ms. Each shot gather includes 320 traces and 560 time samples. See \cite{billette2005} for more information about this dataset.
	
	We design a U-net with convolutional blocks and skip connections following \cite{ronneberger2015, isola2017, abedi2022}.  The skip connections facilitate the flow of low-level information, avoid information loss during the downsampling process, and stabilize the training process by mitigating the vanishing/exploding gradient problem. Table \ref{tab:table1} shows the details of the designed network that includes seven convolutional blocks. 
	
	\begin{table}
		\centering
		\begin{threeparttable}
			
			\caption{Detailed architecture of the designed U-net, shown in Figure \ref{fig2} . \label{tab:table1}}
			\begin{tabular}{lllll}
				\hline
				Block 1	&Conv (4$ \times $4, 2$\times$2, 32), ReLU\\
				Block 2&Conv (3$\times$3, 2$\times$2, 128), BN, ReLU\\
				Block 3	&Conv (3$\times$3, 2$\times$2, 256), BN, ReLU\\
				Block 4	&TrConv (3$\times$3, 2$\times$2, 128), BN, ReLU\\
				Block 5	&TrConv (3$\times$3, 2$\times$2, 32), BN, ReLU\\
				Block 6	&TrConv (3$\times$3, 2$\times$2, 32), BN, ReLU\\
				Block 7	&Conv (3$\times$3, 1$\times$1, 1)\\
				\hline
			\end{tabular}
			\begin{tablenotes}
				\item The network blocks are composed of convolutional (Conv), transposed convolution (TrConv), batch normalization (BN), and concatenation layers, in addition to the rectified linear unit  (ReLU) activation function. In parenthesis, we show the kernel size, stride size, and the number of filters, respectively. Output of each block is the input of the next block. Shortcuts also connect the output of Block 1 to Block 6, and Block 2 to Block 5.
			\end{tablenotes}
		\end{threeparttable}
	\end{table}
	
	Figure \ref{fig1}  shows an example of using the conventional blind-trace denoising method on the synthetic data set. For training, we use 50 instances of each data sample with one randomly selected masked (and active) trace in each one of them. Figure \ref{fig1}a  shows a denoised data sample and its difference with the noisy and the clean data; Figure \ref{fig1}b  shows the average errors of each trace in the denoised data sample, and Figure \ref{fig1}c  shows the evolution of the loss (black curve) and two metrics during training. The metrics are the loss in nonactive traces (calculated by replacing $ \textbf{M} $ with $ 1-\textbf{M }$ in  equation 1), and the difference between the DNN output and the clean data. The original method has generally good performance in reconstructing individual noisy traces from the neighboring clean traces. However, a noisy trace degrades the reconstruction accuracy of its neighboring traces because all traces are reconstructed from their neighbors. Besides, we observe signal leakage in the entire data because of ignoring the signal in each trace during its reconstruction. In Figure \ref{fig1}b, the highest errors are at traces located next to a noisy trace (marked with red circles). A similar performance can be seen in figure 3 from \cite{liu2022coherent}. We alleviate these issues by the method proposed in the following section.
	
	\begin{figure*}[]
		\centering
		\begin{tabular}{@{}p{\linewidth}@{\quad}p{\linewidth}@{}}
			\centering
			\subfigimg[width=.8\linewidth]{(a)}{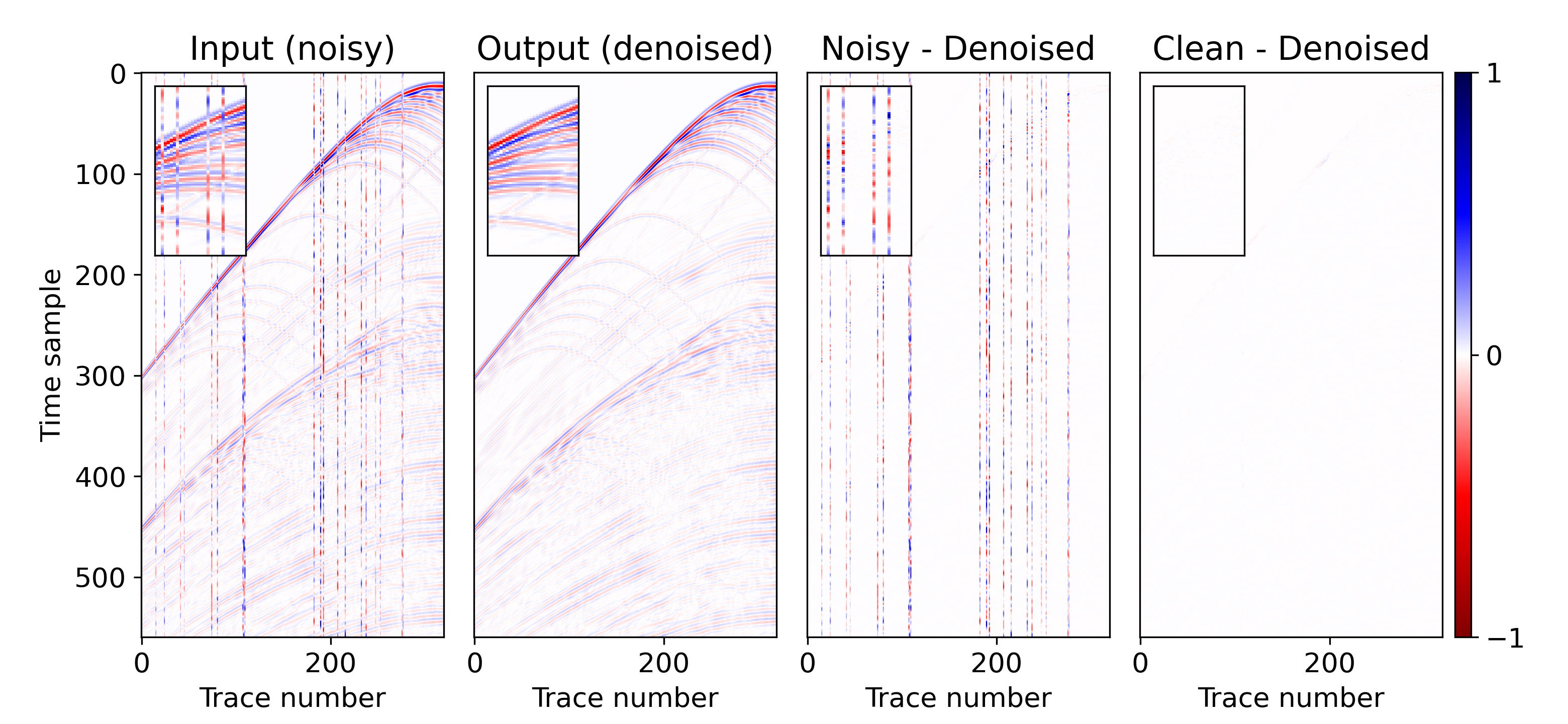} 
			\\
			\subfigimg[width=.45\linewidth]{(b)}{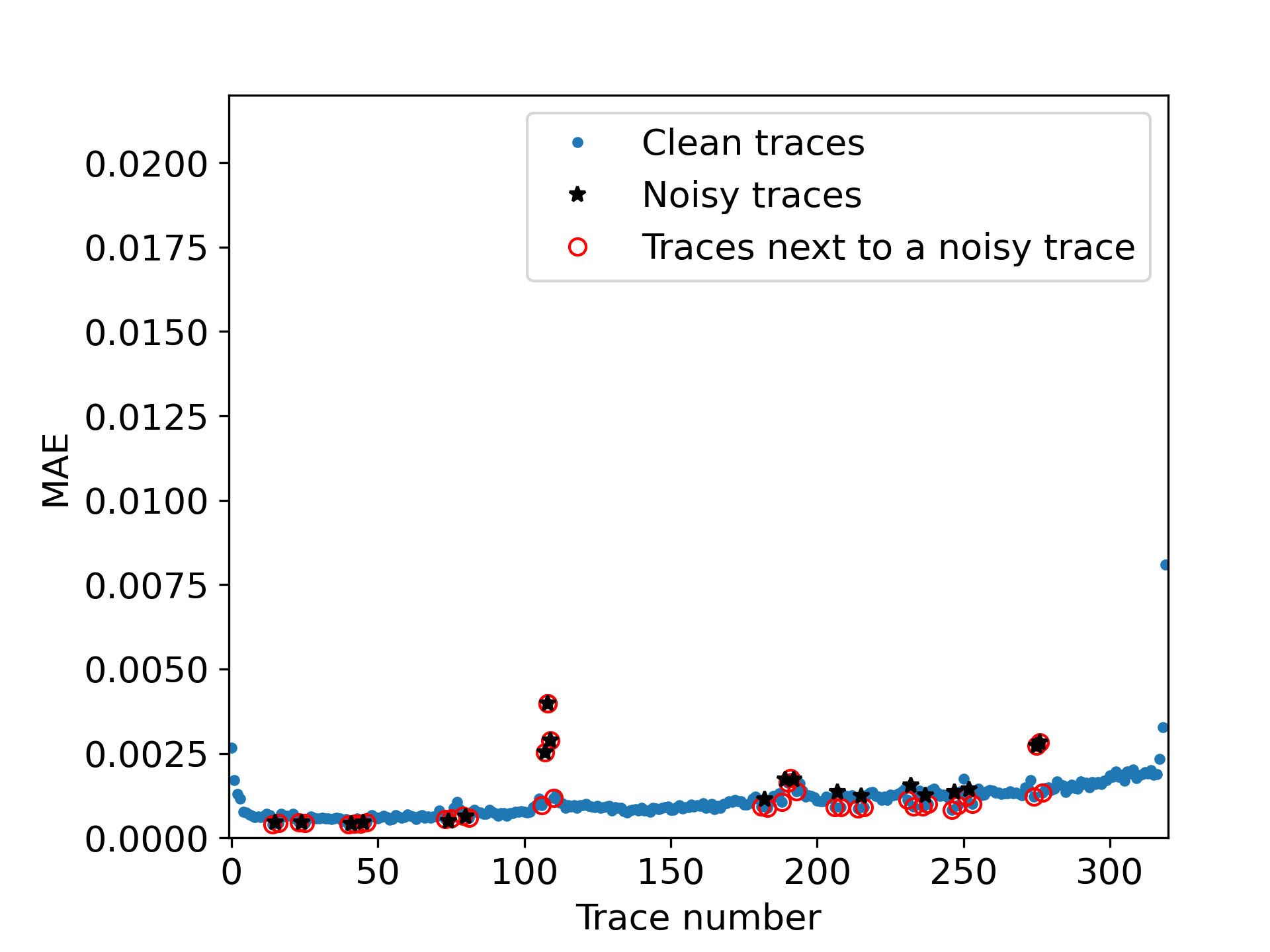} 
			\subfigimg[width=.45\linewidth]{(c)}{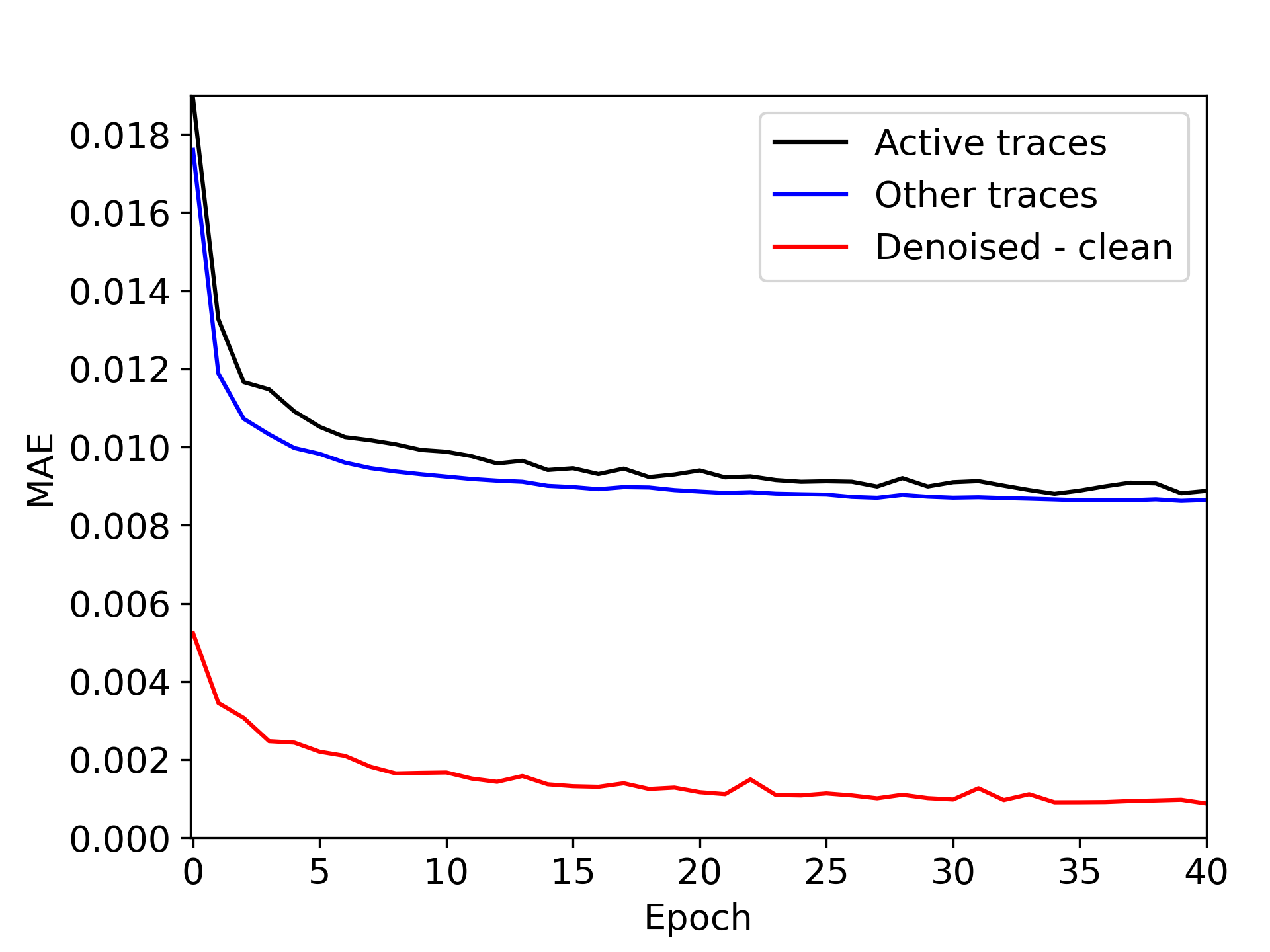}
		\end{tabular}
		\caption{Application of the proposed semi-blind-trace denoising on synthetic data. a) From left to right: the input noisy data, the self-supervised denoised version, the removed noise, and the signal leakage. Overlapping windows show a zoomed portion. b) The average error of each trace after denoising. c) Evolution of loss (in black), and the two metrics explained in Figure \ref{fig1}. To be compared with Figure \ref{fig1}. }
		\label{fig3}
	\end{figure*}

	\section{Semi-blind-trace algorithm with hybrid loss}
	
	Starting from the blind-trace denoising method, herein, we modify the masking process and the definition of active traces in the calculation of the loss function. 
	
	In blind-trace network  \cite{wang2022}, the masking process is embedded in the architecture through a series of padding and cropping. That network performs the masking for each trace in an image separately, processes all instances of the masked image simultaneously, and activates only the masked traces for the calculation of the loss.  Here, we use a more flexible masking procedure with an adjustable number of masked traces, an adjustable masking pattern, and an adjustable definition of active zones. We form a masking matrix $ \textbf{M}_1 $ by placing ones over $ n $ randomly selected traces and zeros elsewhere. The masking is performed through an elementwise matrix multiplication at the input of a simple U-net,
	
	\begin{dmath}
		\label{eq2}
		\textbf{D}_m=\textbf{M}_1 \odot \textbf{B}+(1-\textbf{M}_1 )\odot \textbf{D},
	\end{dmath}
	\noindent
	where $ \textbf{B} $ is the noise matrix with the same dimensions and within the same range of amplitudes as $ \textbf{D} $. An example of the masking process is shown in Figure \ref{fig2}. 
	
	To decrease the signal leakage, we propose to modify the loss function by adding a contribution from the two traces neighboring each masked trace. These traces are the ones showing the maximum error in the original method (Figure \ref{fig1}). This makes the algorithm semi-blind, as it is blinded to the masked traces but not to their neighboring traces. In doing so, the network is trained to reconstruct the noisy traces (ignoring the data in those traces) and reproduce the clean traces (using their data). Therefore, it learns to recognize the vertical trace-wise noise while performing the reconstruction. The modified loss function reads:
	
	\begin{dmath}
		\label{eq3}
		L = \frac{1}{ \sum \textbf{M}_2} \sum \textbf{M}_2 \odot \bigg |\textbf{D} - f ( \textbf{D}_m)\bigg |,
	\end{dmath}
	\noindent
	where the matrix $\textbf{M}_2$ consists of ones over the same traces as $\textbf{M}_1$, a small taper $\varepsilon$ at traces next to them, and zeros elsewhere ($ \textbf{M}_2  $ is not separately random, it is related to $ \textbf{M}_1 $). An example of the calculation of loss using $\textbf{M}_2$ is shown in Figure \ref{fig2}. 
	
	The proposed loss function in equation 3 has active traces over both masked and original (non-masked) traces. The parameter $\varepsilon$ in $\textbf{M}_2$ determines the degree of contribution from the original traces to the loss function. Since in $ \textbf{M}_1 $, for each masked trace there are two non-masked traces, when $\varepsilon = 0.5$, the masked and original traces equally contribute to the loss. Considering that the main objective is to reconstruct the masked traces, the value of $\varepsilon$ should be below 0.5 to avoid a situation where mapping of the original traces dominates the loss. 
	
	We define the number of traces that are masked simultaneously in each sample as $ n $. The use of multiple masked (and active) traces in each training step enables the network to learn the predictability of the signals over a wider range and in the presence of other noisy traces.  Having instances of two (or more) masked (and active) traces located next to each other during training also helps the algorithm in denoising the neighboring noisy traces in the original data. Masking $ n $ randomly selected traces in each data sample, the peak probability is to have $(n-1)/n_{traces}$  masked traces next to each other (i.e. paired). In our example, for $ n=33 $, the mean probability of having paired traces is 10\%  with a standard deviation of ±5\%,  and for $ n=65 $, it is 20\%  with a standard deviation of ±4.4\%.

	\subsection{Implementation}
	
	Figure \ref{fig2} depicts the designed network, the masking process, and the calculation of loss for the proposed semi-blind-trace algorithm. The details of the network architecture are presented in Table 1. We only have the original noisy dataset $ \textbf{D} $, no clean version, and no prior information about the trace-wise noise. To train the network, we follow these steps:
	\begin{enumerate}
		\item{For a mini-batch of the dataset ($ \textbf{D} $), we generate uniform random noise with the same dimensions and within the same range of amplitudes as $ \textbf{D} $, and apply a band-pass filter on it with a randomly selected band to obtain $ \textbf{B} $.}
		\item{We randomly select $ n $ traces and form the matrix $\textbf{M}_1$.  }
		\item{The masked data are then obtained from the original noisy data using $ \textbf{D}_m=\textbf{M}_1 \textbf{B}+(1-\textbf{M}_1 )\textbf{D}$, and are used as the input of our DNN.}
		\item{Forming the matrix $\textbf{M}_2$ based on  $\textbf{M}_1$ and the user-defined value of taper ($\varepsilon$), the loss function is calculated as the sum of the absolute difference between the DNN output $ f( \textbf{D}_m) $ and the original data $ \textbf{D} $, multiplied by $\textbf{M}_2$ (elementwise). }
		\item{The masked minibatch of data is fed to the network and a gradient descent algorithm is applied to form one training step.}
		\item{We repeat steps 2 to 5 several times on the same mini-batch, each time randomly reselecting the $ n $ masked and active traces.}
		\item{Applying step 6 on all mini-batches, we call it one epoch of training.}
	\end{enumerate}
	A Python implementation of the proposed algorithm is available through the link shared in the conclusion section.
	
	Note that $ \textbf{M}_1 $ and $ \textbf{M}_2 $ are related and vary per training step. In one epoch, each data sample is masked and fed to the network several times, using different masks (step 6). Using 50 instances of different masks for each mini-batch in each epoch,  Figure \ref{fig3}a shows the result of applying the proposed method on the same data as in Figure \ref{fig1}. Comparing Figure \ref{fig3}b to Figure \ref{fig1}b, the reconstruction accuracy is improved --especially for paired noisy traces--, and the signal leakage is reduced --especially at traces next to a noisy trace. Comparing Figures \ref{fig3}c and  \ref{fig1}c, the presented metric shows a reduction in the errors in the denoised data. 
	
	In step 1, we randomly change the frequency content of the noise data in masks from mono frequencies to a wide range of frequencies. If we expect to have noise that partly covers the traces (e.g., the swell, spike-like, and velocity sensor noise in marine data \cite{hlebnikov2021}) we can randomly change the temporal location of masks to teach the network to look for horizontal incoherent noise in smaller parts of a trace instead of the entire trace.
	
\begin{figure*}[]
		\centering
		\begin{tabular}{@{}p{\linewidth}@{\quad}p{\linewidth}@{}}
			\centering
			\subfigimg[width=.9\linewidth]{(a)}{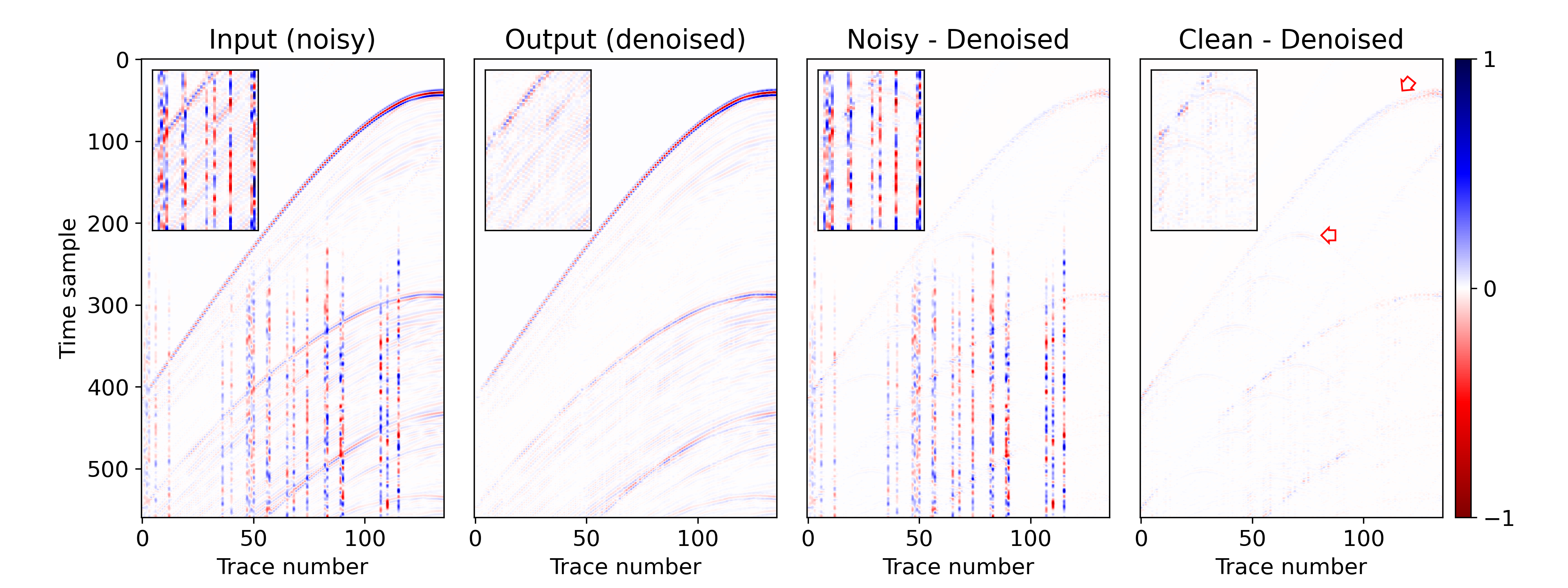} 
			\\
			\subfigimg[width=.9\linewidth]{(b)}{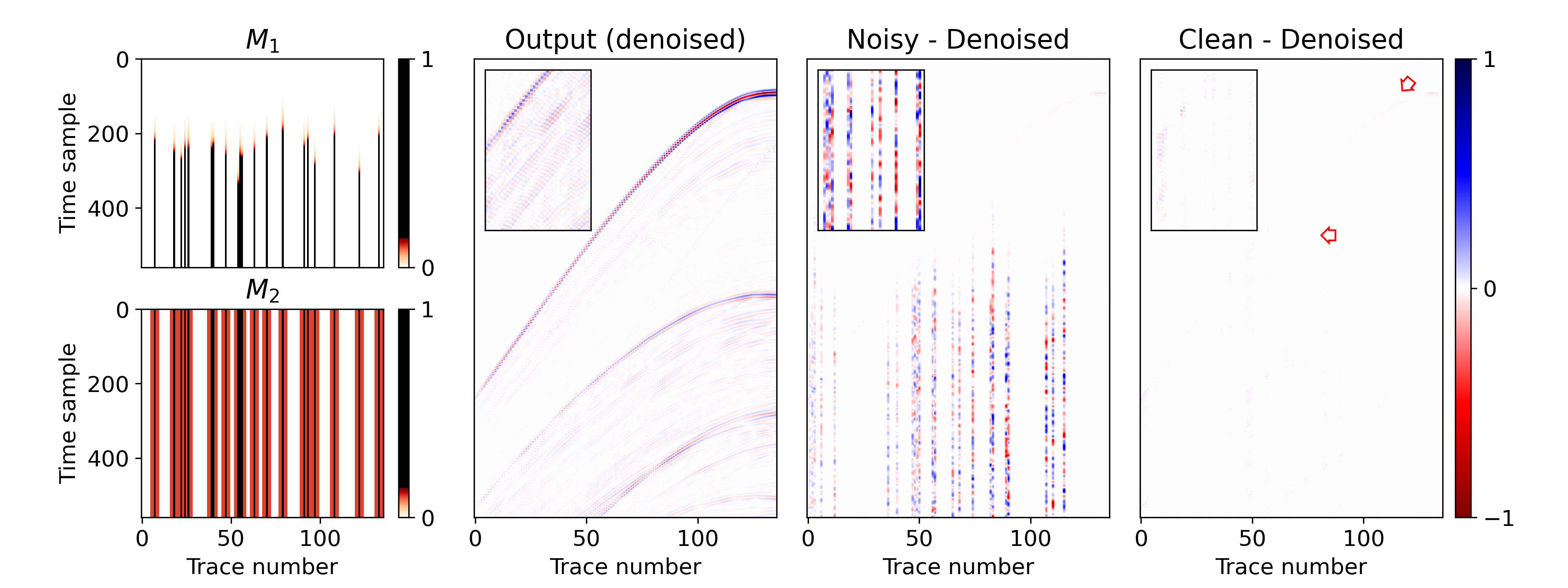} 
		\end{tabular}
		\caption{Application of the original blind-trace (a), and the proposed semi-blind-trace denoising on synthetic data with unknown partially noisy traces. In (a) we see signal leakage in originally clean events (marked by red arrows). In (b) the signal leakage is prevented after adjusting the masking matrices for the objective noise pattern. Note the denoising of four noisy traces located next to each other.}
		\label{fig4}
	\end{figure*}

	As an example of attenuations of a different noise pattern, we create synthetic data with vertically coherent noise that partly covers an unknown portion of the traces (Figure \ref{fig4}).  Initially, we employ the conventional blind-trace denoising, as explained in section 2. In Figure \ref{fig4}a, besides the denoising errors, we see signal leakage in events that are originally clean (marked by red arrows).  Next, we employ the proposed method using masking matrices ($ \textbf{M}_1 $) that are adjusted to the objective noise pattern (Figure \ref{fig4}b). We also modify $ \textbf{M}_2 $ to include two active traces on each side of a masked trace. This helps to reduce signal leakage when the trace interval is large. The combination of the adjusted masking pattern and active zones enables the network to learn that the objective noise partly covers the traces. Compared to Figure \ref{fig4}a, the signal leakage of the conventional method is prevented in Figure \ref{fig4}b.
	
	The proposed method has two hyperparameters, $ n $ and $\varepsilon$. In the following, we study their role and show how to appropriately select them before training. 
	
	\section{Numerical analysis}
	
\begin{figure}[]
		\centering
		\includegraphics[width=.5\linewidth]{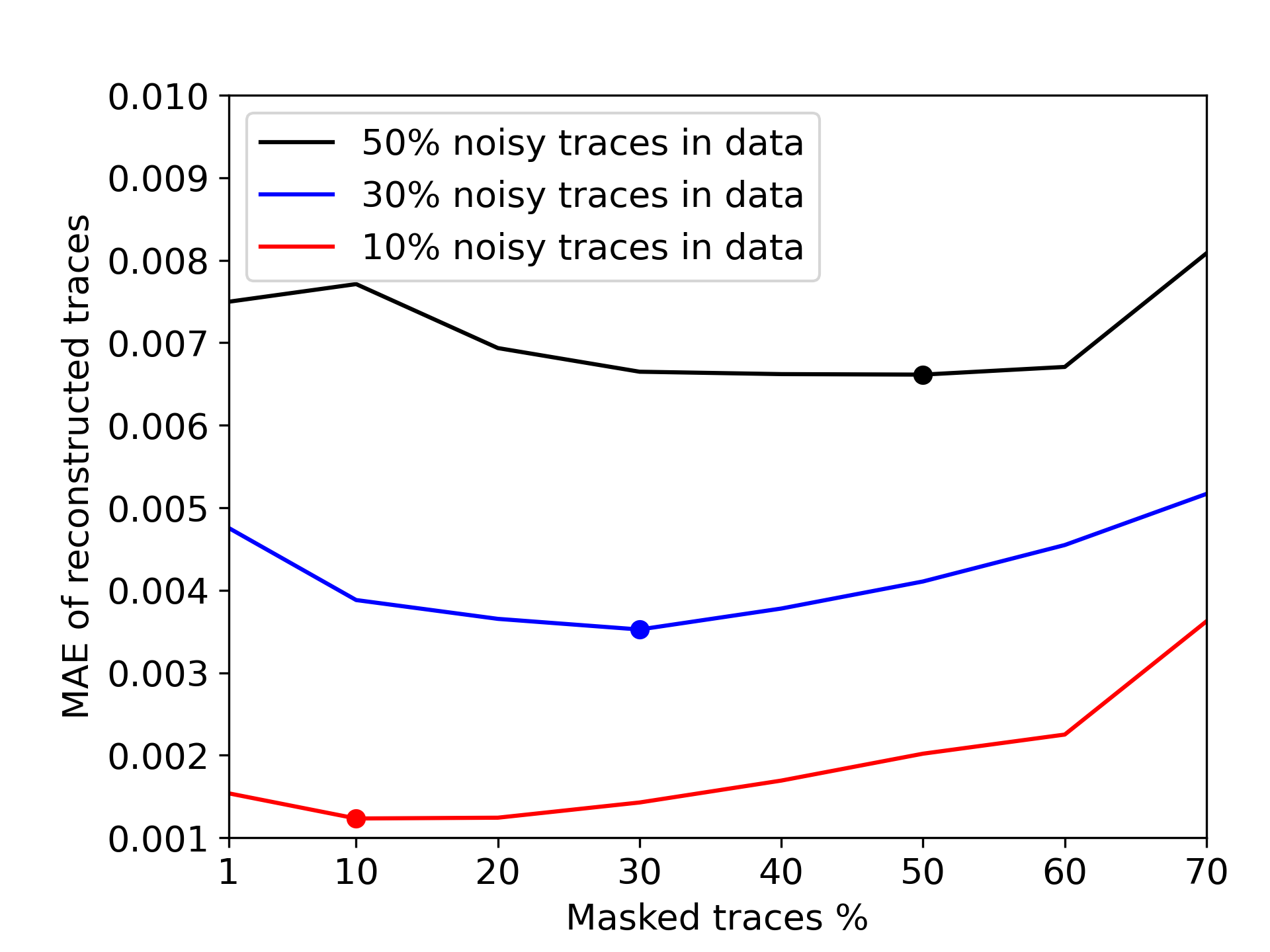}
		\caption{Studying the effect of the number of masked traces ($ n $) on three datasets with different portions of noisy traces. Our method works best when $ n $ is selected close to the number of noisy traces in the original data.}
		\label{fig5}
	\end{figure}

	The parameter $ n $ determines the number of masked and active traces during training (the number of one-valued traces in $ \textbf{M}_1 $ and $ \textbf{M}_2 $). We design a numerical experiment to study its effect on the proposed method. We use the synthetic data considered in the blind-trace denoising section and shown in Figures \ref{fig1} and \ref{fig3}. We create three separate noisy training datasets by replacing 10\%, 30\%, and 50\% randomly selected traces in each shot-gather with band-limited noise.  We use the proposed method for each dataset and train the DNN using different values of $ n $ corresponding to 1\% to 70\% of the number of traces. Figure \ref{fig5} shows the best performance of the network for each scenario. When the selected value of $ n $ is close to the number of noisy traces in the data that we want to denoise, the noisy traces are reconstructed with higher accuracy. 
	
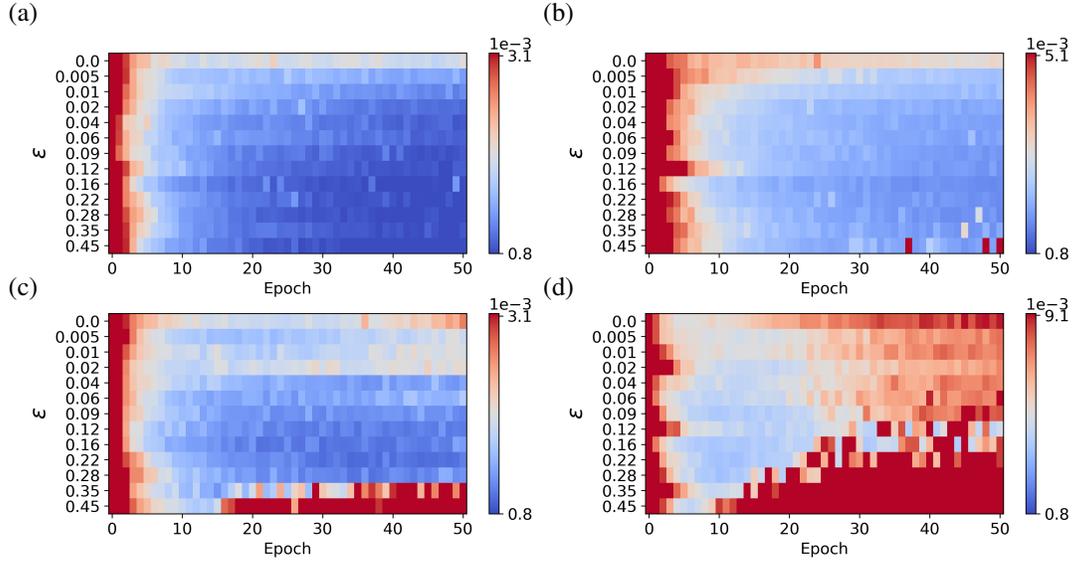
\begin{figure*}[]
		\centering
		\begin{tikzpicture}
			\node[anchor=south west, inner sep=0] (image) at (0,0) {\includegraphics[width=.9\linewidth]{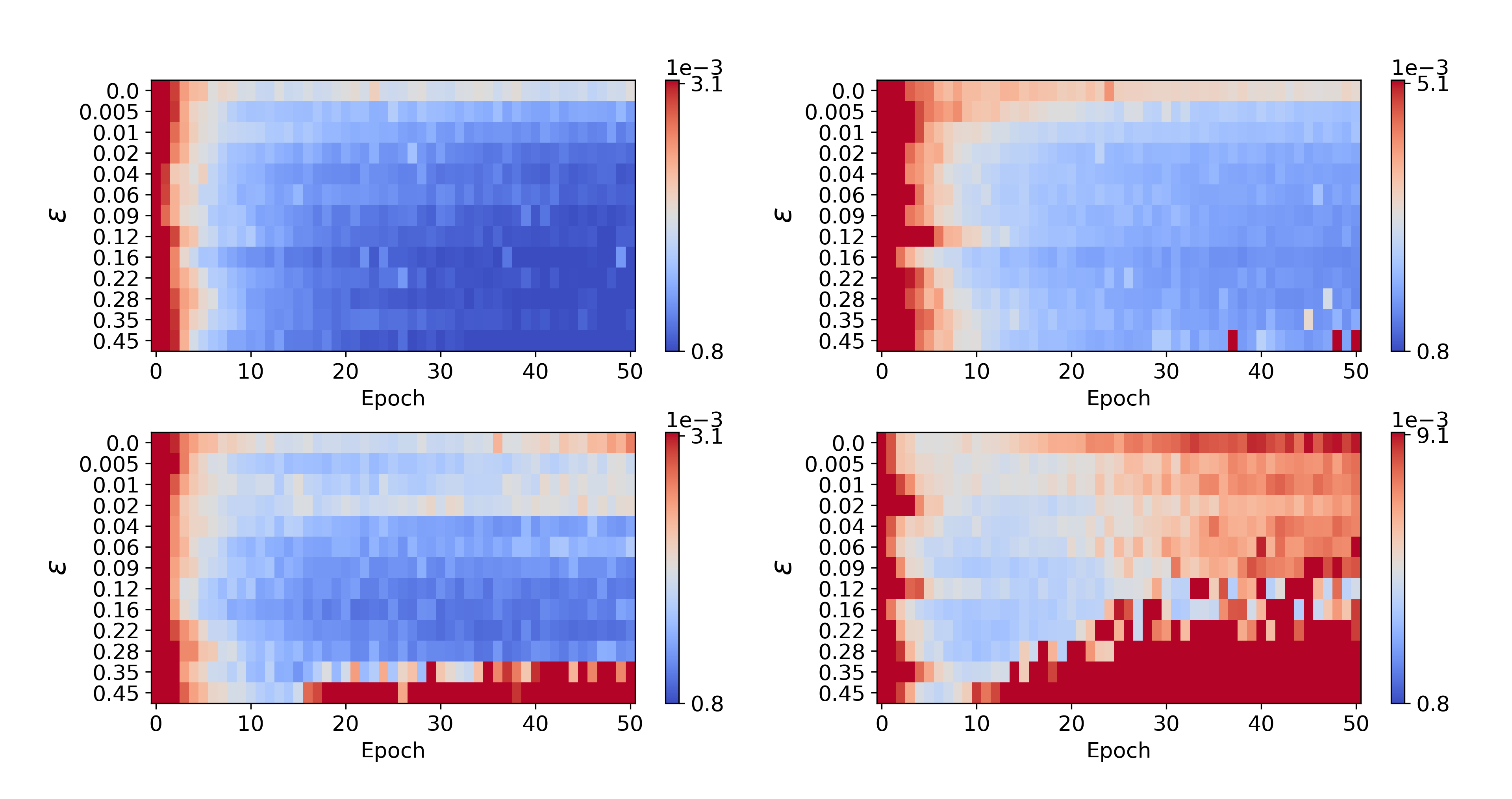}};
			\node[anchor=north west, xshift=1pt] at (image.north west) {(a)};
			\node[anchor=north, , xshift=1pt] at (image.north) {(b)};
			\node[anchor=west, yshift=0.5\pgflinewidth , xshift=1pt] at (image.west) {(c)};
			\node[anchor=center, yshift=0.5\pgflinewidth, , xshift=1pt] at (image.center) {(d)};
		\end{tikzpicture}
		\caption{Studying the effect of $\varepsilon$ (determines the contribution of non-masked traces in the loss function) and noise pattern on denoising accuracy. We show the mean absolute error of the DNN output as a function of $\varepsilon$ (vertical axis) and training epochs (horizontal axis) for datasets with a) 32 noisy traces with varying positions, b) 96 noisy traces with varying positions, c) 32 noisy traces with fixed positions, d) 96 noisy traces with fixed positions. }
		\label{fig6}
	\end{figure*}

	Parameter $\varepsilon$ in our method determines the degree of contribution of non-masked traces next to a masked trace in the loss function. To study the role of $\varepsilon$, we use the aforementioned datasets with 10\% and 30\% noisy traces. We use the proposed method for each dataset and train the DNN using different values of $\varepsilon$ ranging from 0 to 0.45. Figures \ref{fig6}a and \ref{fig6}b respectively show the error in the DNN output versus the clean data at the end of each training epoch for datasets with 10\% and 30\% noisy traces with varying positions. We observe that even a small value of $\varepsilon$ considerably decreases the errors compared to $\varepsilon=0$. Increasing the selected value for $\varepsilon$ generally decreases the errors. The noisier data also show a higher reconstruction error, as expected. 
	
	In the above tests, the position of the noisy traces varies from one shot gather to another. It can represent the trace-wise noise pattern for external sources, and receiver malfunctioning in land data. Here, we change the noise pattern so that the noisy traces have a (randomly selected) fixed position in all shot gathers. This can represent hydrophone malfunctioning in offshore data. Figures \ref{fig6}c and \ref{fig6}d respectively show the error in the DNN output with respect to the clean data at the end of each training epoch for datasets with 10\% and 30\% noisy traces that have fixed positions. The errors increased compared to Figures \ref{fig6}a and \ref{fig6}b. Besides, we observe that errors initially decrease, then start to increase as the training progresses. Note that these errors are different from the loss. The loss is the error in the DNN output versus the noisy data at active traces, which continues to decrease. This increase in the reconstruction error occurs when the network starts to reproduce the noisy traces in the output (resulting in a fall in the loss), therefore, the training should be stopped. 
	
	From these numerical tests, we conclude that optimal values of both parameters are related to how noisy the original dataset is. We suggest selecting $ n $ to approximate the number of noisy traces that are expected to be in the dataset. the mild variations in the curves in Figures \ref{fig5} suggest, the results are not sensitive to the exact value of  $ n $. If the original data have noisy traces with a fixed position in all datasets, early stopping is crucial to avoid a reproduction of noise in the output. In this case, augmentation methods like shifting and flipping the data are suggested. Selecting a small value for $\varepsilon$ helps to avoid the reproduction of noisy traces in early epochs, while larger values of $\varepsilon$ help to reproduce the clean data more accurately. In our tests, we had the best results for $\varepsilon$ within the range of 0.001 to 0.2.
	
	\section{Application}
	
	\begin{figure*}[]
		\centering
		\includegraphics[width=1\linewidth]{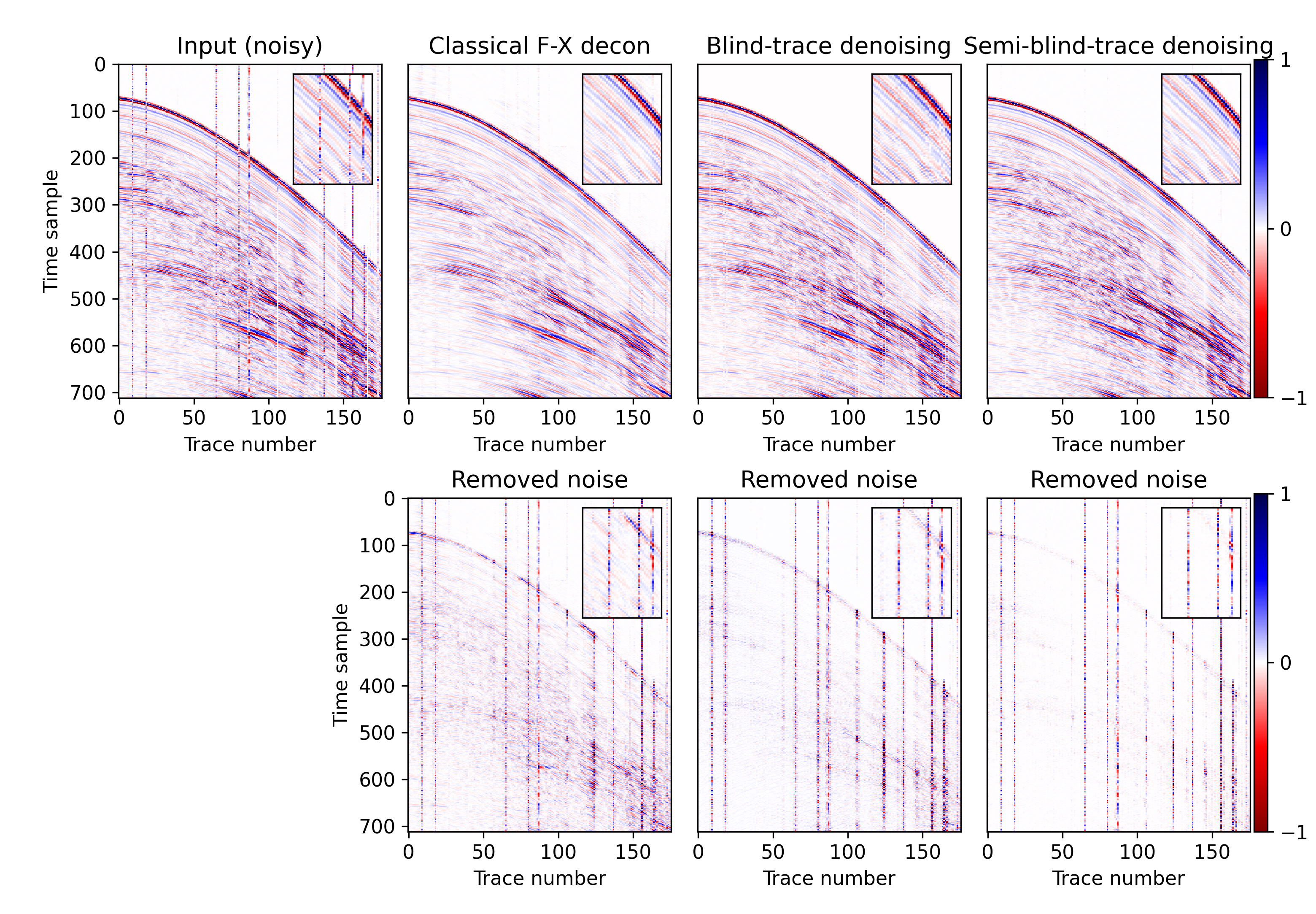}
		\caption{Comparing the proposed semi-blind-trace method with the original blind-trace denoising and f-x devolution filtering on a real marine dataset. The upper row shows the input and the denoised data, and the lower row shows the difference between the input and the denoised data.}
		\label{fig7}
	\end{figure*}
	Here, we show the application of the proposed method on two real datasets that have noisy traces in the raw recorded data. The first dataset consists of 300 shot gathers from vintage offshore data recorded in the Gulf of Mexico. Each shot gather consists of 176 traces with 720 time samples. Figure \ref{fig7} shows a representative shot gather. Each raw shot gather originally has 12 to 17 noisy traces with different noise types that include high- and low-frequency noise, partly zero traces, and a polarity reversal. Most of the noise is due to malfunctioning of the hydrophones, therefore, occurs at fixed trace numbers (fixed offsets) within all shot gathers. We manually add noise to trace number 86 to create an instance of two neighboring noisy traces for this experiment. For data augmentation, we use flipping and horizontal shifting of the data samples that change the position of the noisy traces. We select $ n=15 $ (as a rough estimation of the number of noisy traces) and $\varepsilon=0.15$ (through try and error) and apply the proposed fully self-supervised method following the steps in the implementation section. Note that these parameters do not need careful determination for a suitable result (no visible difference was seen by changing them $\pm 30 \%$). During training, the noise data in the masks are generated using random band-pass filtering ()as explained in the Implementation subsection) without any adjustment for noise characteristics of this dataset.
	
	Figure \ref{fig7} compares the denoising performance of the original blind-trace and the proposed semi-blind-trace deep learning methods alongside the classical f-x prediction filtering \cite{canales1984,gulunay1986fxdecon}. The f-x filter is a traditional method for attenuating horizontally incoherent noise; but shows poor signal reconstruction  \cite{gulunay2017signal,hlebnikov2021,farmani2023}. We applied the f-x deconvolution in small overlapping windows and then averaged them for a better result.
	
	Since most of the noisy traces have fixed trace numbers in all shot gathers, early stopping is necessary to avoid the regeneration of noisy traces in the output, as discussed in the Numerical analysis section. We stop the training after 25 epochs for the original method and 20 epochs for our proposed method. The early stopping and incoherencies that exists in real raw data result in the remainder of energy in all traces that pass the U-net (Figure \ref{fig7}). The proposed method reconstructs the noisy traces with the highest accuracy, and the other traces are reproduced with the lowest signal leakage among the compared methods.
	
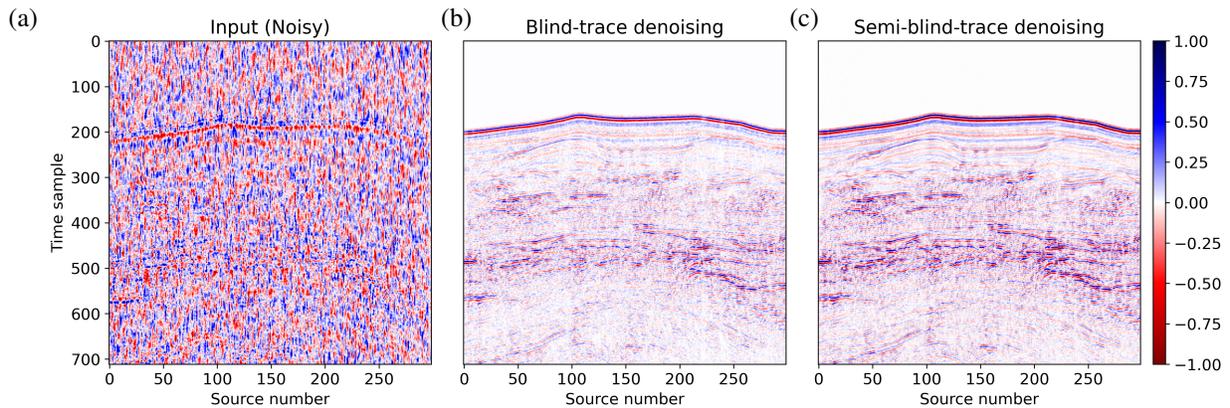
\begin{figure*}
		\centering
		\begin{tikzpicture}
			\node[anchor=south west, inner sep=0] (image) at (0,0) {\includegraphics[width=1\linewidth]{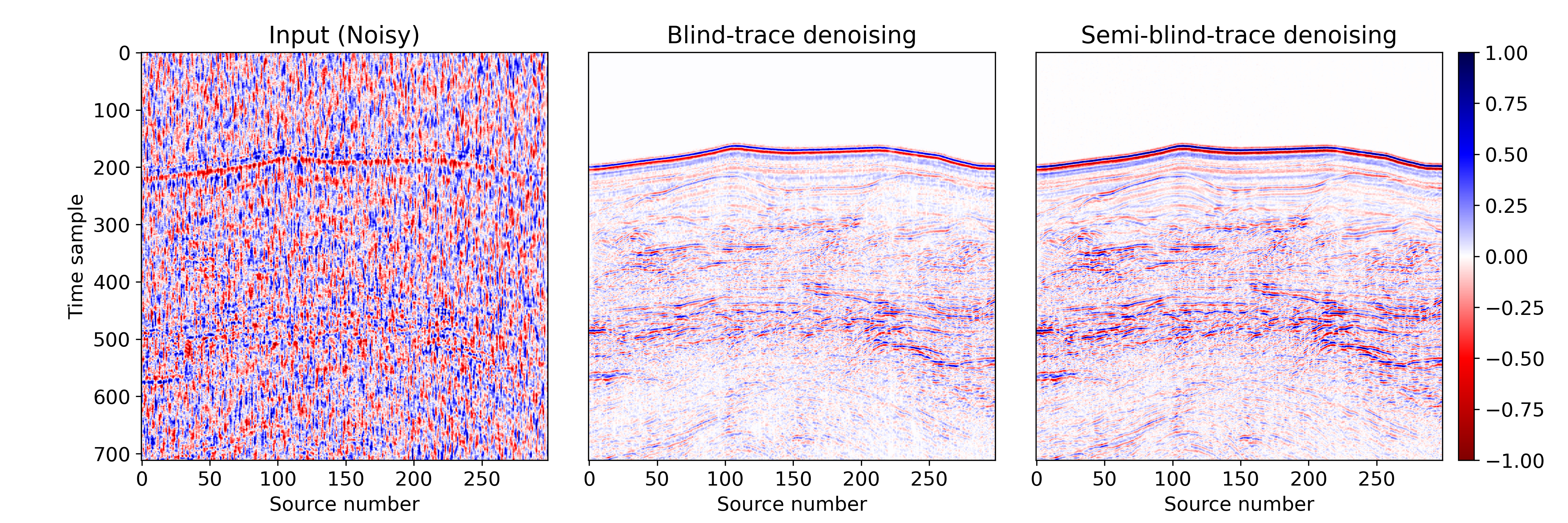}};
			\node[anchor=north west, xshift=1pt] at (image.north west) {(a)};
			\node[anchor=north, xshift=-0.13\linewidth] at (image.north) {(b)};
			\node[anchor=north, xshift=0.15\linewidth] at (image.north) {(c)};
		\end{tikzpicture}
		\caption{A common-offset-gather from hydrophone number 87 in Figure \ref{fig7}. a) The noisy data, b) gathered from denoised shot gathers with original blind-trace deep learning, c) gathered from denoised shot gathers with our semi-blind-trace algorithm.}
		\label{fig8}
	\end{figure*}

	Figures \ref{fig8}a, \ref{fig8}b and \ref{fig8}c show common offset gathers from hydrophone number 87 (that is next to another noisy hydrophone), before and after denoising the shot gathers with the original and our deep learning methods. The reconstructed data from different shot gathers are more consistent using our method.

\begin{figure*}
		\centering
		\begin{tikzpicture}
			\node[anchor=south west, inner sep=0] (image) at (0,0) {\includegraphics[width=1\linewidth]{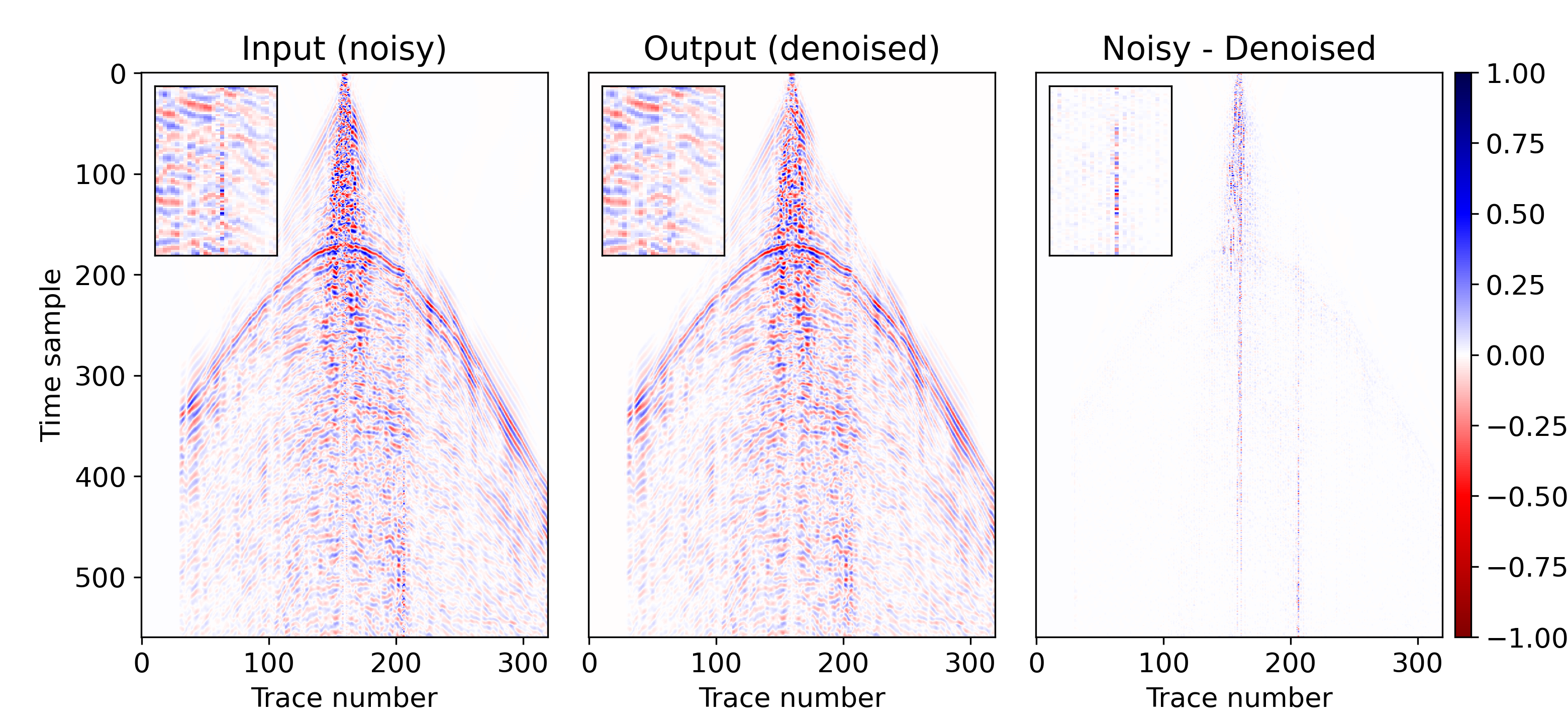}};
			\node[anchor=north west, xshift=1pt] at (image.north west) {(a)};
			\node[anchor=north, xshift=-0.13\linewidth] at (image.north) {(b)};
			\node[anchor=north, xshift=0.15\linewidth] at (image.north) {(c)};
		\end{tikzpicture}
		\caption{Application of the proposed semi-blind-trace method on a land dataset with three highly noisy traces in the recorded data. We display an original shot gather, the denoised version, and the difference between them.}
		\label{fig9}
	\end{figure*}
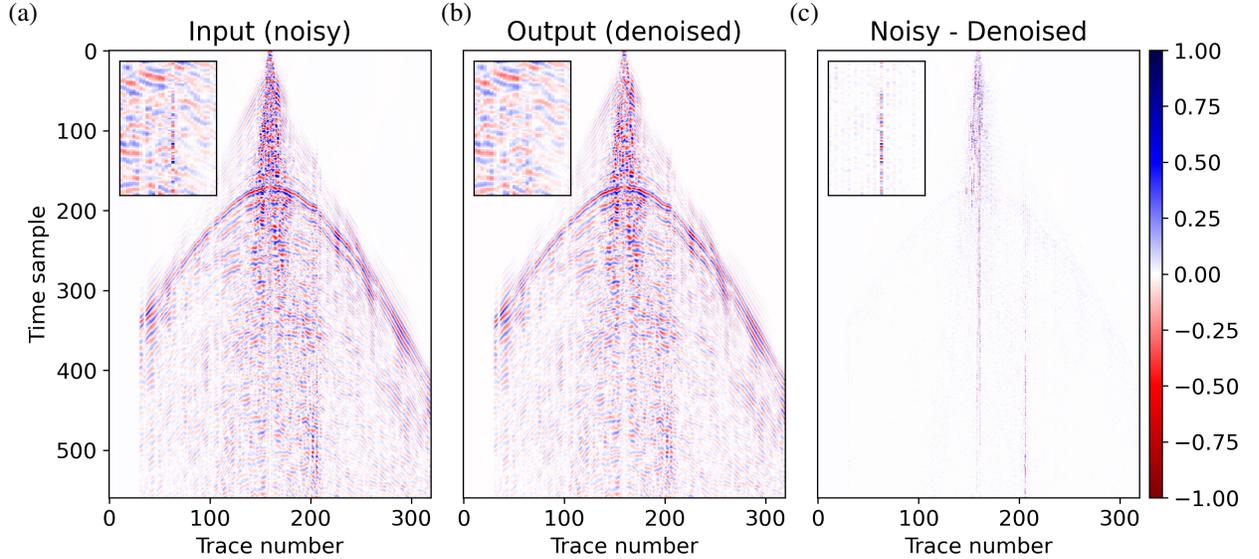

	The second dataset that we use is land data recorded with vibriosis using a fixed shot and receiver interval. We have 205 shot gathers, each having 320 traces recording 560 time samples. The noisy traces that we want to denoise are recorded by one malfunctioning receiver, and the two receivers closest to the source. These trace-wise noises are shown in a representative example of the data in Figure \ref{fig9}a. As a data augmentation, we add horizontally flipped samples to the dataset. We select $ n=20 $ and $\varepsilon=0.1$ and apply the proposed method following the steps in the implementation section. Figure \ref{fig9}b shows the denoised version that our trained DNN produces, and Figure \ref{fig9}c shows the difference between the original and denoised shot gather. Only the vertical noises are attenuated, and all the other events are reproduced as the original data. 
	
	In the three examples that we presented here, we trained the network with our proposed method for respectively 30, 25, and 20 epochs to reach the desired reconstruction accuracy. Using a GPU (GV100 with 32GB VRAM), the training times were respectively 11, 18, and 9 minutes. However, the training time of a deep learning method depends on many factors like the data size, batch size, coding method, and machinery capabilities.
	
	\section{Discussion}
	Many existing denoising methods need a synthetic generation of accurately representative noises. Considering that the source and different aspects of different noises are unknown, we aim to train a network to recognize and remove trace-wise noise without any knowledge about the frequency range or the location of the noise. 
	
	We proposed two major modifications to existing blind-trace methods. First, we create active zones around each masked trace that is larger than the masked area. This enables the network to distinguish between horizontally predictable signals and unpredictable noise. Second, to force this predictability as the main distinguishable characteristic, we dynamically change the characteristics and location of masks in each training step. Since the noise in the masks is constantly changed, while the signal data are fixed, the network is trained to learn the characteristics of the signals instead of the noise. A similar strategy may also be implemented for other types of noise with other unknown characteristics.
	
	Our method is also capable of reconstructing several incoherent noisy traces next to each other because such cases may happen during randomly masking several traces.  Attenuating noises with horizontal coherency needs further investigation on the required modifications of mask shapes and characteristics.
	
	\section{Conclusion}
	We proposed a flexible semi-blind-trace self-supervised algorithm to attenuate laterally incoherent noise in seismic data. The proposed masking procedure is a fast and online process that is adjustable to the noise pattern and allows a network to learn the reconstruction of neighboring noisy traces. We use a hybrid loss function that calculates a weighted error between the DNN output and the input noisy data at the masked traces and their adjacent non-masked traces. Compared to conventional blind-trace denoising, our method does not blindly reconstruct all traces from their surrounding traces; it learns to detect the noisy traces, reconstruct them from the clean data, and pass the clean data to the output. Using the proposed training steps, we observe a considerable reduction in the effect of noisy traces on the reconstruction of nearby traces, and an enhancement in signal fidelity. We have two hyperparameters: the number of masked traces in each data sample ($ n $), and the relative weight of masked and non-masked traces in the loss function ($\varepsilon$). The numerical experiments show that $ n $ should approximate the number of noisy traces in the data to have more accurate denoising. The higher the $ \varepsilon $, the lower the signal leakage, but also the higher the risk of noise reproduction in early epochs especially when the noisy traces have a fixed position in all samples. We saw considerable improvements in signal leakage even with small values of $ \varepsilon $. Therefore, we suggest selecting a small value for $\varepsilon$. It can be increased if the signal leakage exceeds the desired criterion. The presented applications on two field datasets showed the robustness of the proposed method in attenuating trace-wise noise while preserving the data fidelity. A Python implementation of the proposed algorithm and a demo example is available at \url {https://github.com/mahdiabedi/semi-blind-trace-deep-learning}.
	
	\section*{Acknowledgments}
	This work has received funding from: the Spanish Ministry of Science and Innovation projects with references TED2021-132783B-I00, PID2019-108111RB-I00 (FEDER/AEI) and PDC2021-121093-I00 (MCIN / AEI / 10.13039 / 501100011033 / Next Generation EU), the “BCAM Severo Ochoa” accreditation of excellence CEX2021-001142-S / MICIN / AEI / 10.13039 / 501100011033; the Spanish Ministry of Economic and Digital Transformation with Misiones Project IA4TES (MIA.2021.M04.008 / Next Generation EU PRTR); and the Basque Government through the BERC 2022-2025 program, the Elkartek project SIGZE (KK-2021 / 00095), and the Consolidated Research Group MATHMODE (IT1456-22) given by the Department of Education. Tariq Alkhalifah thanks KAUST for its support. We acknowledge BP for making the 2D synthetic data available.
	
	\section*{Data Availability}
	Data in this study cannot be shared.
	
	The authors report no conflict of interest.
	
	\bibliography{references_denoise}

\end{document}